\documentclass[%
 reprint,
superscriptaddress,
 amsmath,amssymb,
 aps,
prb,
]{revtex4-2}

\usepackage{graphicx}
\usepackage{dcolumn}
\usepackage{bm}
\usepackage{xcolor}
\usepackage{soul}

\usepackage[colorlinks=true, linkcolor=blue, citecolor=blue, urlcolor=blue]{hyperref}

\begin{document}

\preprint{APS/123-QED}

\title{Bright yet dark: how strong coupling quenches exciton-polariton radiation} 

\author{Jiaxun Song}
\thanks{These authors contributed equally}
\affiliation{%
 Department of Physics and Astronomy, University of Pennsylvania, Philadelphia, PA 19104, USA
}
\author{Li He}
\email{li.he001@montana.edu}
\thanks{These authors contributed equally}
\affiliation{%
 Department of Physics, Montana State University, Bozeman, MT 59717, USA
}
\author{Bo Zhen}%
\email{bozhen@sas.upenn.edu} 
\affiliation{%
 Department of Physics and Astronomy, University of Pennsylvania, Philadelphia, PA 19104, USA
}%




\date{\today}

\begin{abstract}

Exciton-polaritons offer a promising platform for nonlinear and quantum photonics, but their short radiative lifetimes severely limit the buildup of interaction-driven phenomena.
Understanding and controlling polariton radiative decay is therefore essential for accessing new regimes of polaritonic physics. 
Conventional coupled oscillator descriptions typically treat polariton radiation in terms of the independent radiative decay of bare excitonic and photonic resonances, neglecting how strong exciton-photon coupling can fundamentally reshape their radiative properties.
Here, we develop a general theory of polariton radiation that explicitly captures the collective and coherent nature of exciton-photon coupling.
We show that this coupling can strongly suppress polariton radiation via destructive interference, both within the excitonic ensemble and between excitonic and photonic radiation channels, leading to polaritonic bound states in the continuum with infinitely long radiative lifetimes.
Our approach provides a unified description of polariton radiative decay and establishes new design principles for engineering long-lived exciton-polaritons with tailored radiation properties, opening opportunities for nonlinear, topological, and quantum polaritonic phenomena.

\end{abstract}

\maketitle



Exciton-polaritons are hybrid quasiparticles formed through the strong coupling between excitonic resonances in solid-state materials and confined optical modes. 
By inheriting interactions from their matter component while retaining the coherence and optical accessibility of photons, polaritons provide a promising route to strongly interacting photonic systems. 
This dual light-matter nature has enabled a rich landscape of collective and nonlinear phenomena, ranging from polaritonic Bose–Einstein condensation \cite{byrnes2014exciton} to polaritonic circuitry \cite{sanvitto2016road}. 
The recent integration of two-dimensional (2D) quantum materials, such as transition metal dichalcogenides (TMDs), into engineered nanophotonic structures has further expanded this landscape by combining strong 2D excitonic resonances with nanoscale optical confinement, creating a particularly promising platform for strongly interacting and nonlinear polariton physics \cite{zhang2021van,genco2025femtosecond,wang2026strongly}. 
Yet, the same strong radiative coupling that enables efficient light-matter hybridization also imposes a fundamental limitation: polaritons rapidly lose energy to free space through both their excitonic and photonic components. 
This limitation is especially severe in 2D exciton-polariton systems, where the large exciton oscillator strength enhances radiative decay \cite{wang2018colloquium}, often restricting polariton lifetimes to below $\sim$ 1 ps.
These lifetimes can be comparable to or even shorter than the timescales required for polariton interactions or thermalization, strongly constraining access to collective and nonlinear regimes \cite{byrnes2014exciton}. 
Extending polariton lifetimes while retaining strong light-matter coupling therefore represents a central challenge for realizing the full potential of 2D polaritonic systems.

Conventionally, exciton-polariton systems are described by a non-Hermitian coupled oscillator model \cite{deng2010exciton,zhang2018photonic,chen2020metasurface,khestanova2024electrostatic}, in which the bare exciton and photon resonances are each assigned a fixed, independent radiative decay rate.
Strong hybridization between these two resonances yields polariton eigenstates with complex energies, whose real and imaginary parts give the eigenenergies and decay rates, respectively.
While this phenomenological treatment correctly describes the polariton dispersion, it reduces the radiative decay rate to a simple weighted average of the bare excitonic and photonic components, thereby overlooking two microscopic interference mechanisms rooted in strong exciton-photon coupling that fundamentally reshape polariton radiation.
First, in nanophotonic structures, the characteristic length scale of a photonic mode ($\geq$ hundreds of nm) is far larger than the exciton Bohr radius ($\sim$1nm). 
As a result, the photonic mode couples to a dense ensemble of excitonic modes distributed across this length scale, in a manner reminiscent of Dicke superradiance \cite{dicke1954coherence}. 
These excitonic modes combine coherently to form a single collective exciton mode that alone hybridizes with the photonic mode. 
Crucially, the spatial profile of this collective exciton mode mirrors that of the optical field, which can vary rapidly in space, such as in photonic crystal (PhC) slabs \cite{he2023polaritonic}. 
Because of this spatially varying profile, local excitons within the collective state radiate with different phases and amplitudes -- much like elements of an antenna array -- leading to destructive interference that suppresses, and in some cases fully quenches, radiation \cite{dang2022realization,kravtsov2020nonlinear,ardizzone2022polariton,wu2024exciton}.
In contrast, the bare exciton radiative rate used in the coupled oscillator model is typically associated with spatially uniform excitation, under which local excitons are driven in phase and radiate constructively.
Second, because polaritons are coherent superpositions of the collective exciton and photon modes, their emission is governed by the interference between the radiation channels of these underlying constituents. This interference can likewise be fully destructive, producing polariton states with vanishing radiation even though neither constituent mode is individually dark. 
These interference mechanisms are absent from the phenomenological coupled oscillator model, motivating a more rigorous theoretical framework that captures them explicitly.

Here, we develop a general microscopic theory of polariton radiation that goes beyond the coupled oscillator model.
Starting from a Hamiltonian that explicitly incorporates the coupling between excitons, photons, and the radiation continuum, we show how these two interference mechanisms emerge and govern polariton radiative decay. 
To illustrate the consequences of this framework, we apply it to polaritonic bound states in the continuum (BICs) -- exotic polaritonic states with infinite radiative lifetimes that have recently been realized in diverse nanophotonic platforms \cite{dang2022realization,kravtsov2020nonlinear,ardizzone2022polariton,wu2024exciton}.
We show that these two interference mechanisms give rise to two distinct types of polaritonic BICs: one at high-symmetry points in momentum space, where symmetry quenches both the photonic and collective exciton radiation channels, and another at generic momenta, where destructive interference between the excitonic and photonic radiation channels cancels the net radiation. 
These theoretical predictions are corroborated by full-wave electromagnetic simulations using realistic material parameters, confirming that our microscopic theory accurately captures polariton radiation in nanophotonic systems.

\begin{figure}[htbp]
\includegraphics[width=1\columnwidth]{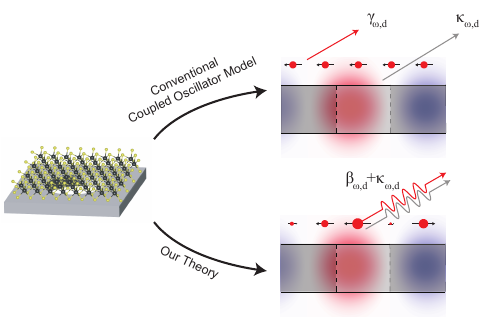}
\caption{\label{fig:pdf1} Visual comparison of radiative mechanisms in coupled TMD-PhC slab. The conventional coupled oscillator model assumes excitons radiate at their intrinsic amplitude $\gamma_{\omega,d}$, which corresponds to an array of uniformly polarized dipoles (represented as red dots with arrows). Radiation from the exciton and photon modes is treated as two independent pathways, represented as two separate, straight lines. 
In contrast, our theory predicts that local exciton modes form a collective state with a modified radiative amplitude $\beta_{\omega,d}$, represented here as non-uniform dipoles whose amplitudes mirror the optical field profile. Radiation into the continuum is visualized as overlapping waves, illustrating the coherent interference between the exciton and photon radiation channels inherent to our theory.   
}
\end{figure}

We begin by considering a MoSe$_2$ monolayer placed in close proximity to a Si$_3$N$_4$ photonic crystal slab ($n=2$) patterned with a square lattice of air holes, as depicted in Fig.~1. 
The PhC slab supports an isolated, dispersive photonic band near the exciton resonance, whereas the exciton band exhibits negligible dispersion and remains flat across the entire photonic Brillouin zone (BZ).
The excitonic and photonic modes strong couple while simultaneously radiating into free space.
Unlike previous formalisms that treat polariton radiative decay phenomenologically by assigning complex energies to the excitonic and photonic modes \cite{zanotti2022theory,zanotti2024legume}, we explicitly incorporate their coupling to the radiation continuum into our Bloch Hamiltonian.
This is essential for capturing interference between excitonic and photonic radiation channels. 
At momentum $\textbf{k}$, this Bloch Hamiltonian can be written as:
\begin{eqnarray}
H_\textbf{k}&=&\hbar \omega_\text{p} a^\dagger_\textbf{k} a_\textbf{k} + \hbar \omega_\text{x} \sum_{\textbf{k}',\sigma} b_{\textbf{k}',\sigma}^\dagger b_{\textbf{k}',\sigma} 
+ \sum_{\omega,d}   \hbar \omega s^\dagger_{\omega,d} s_{\omega,d}  \nonumber \\
    &&+ \sum_{\textbf{k}',\sigma} ( g_{\textbf{k}',\sigma} a^\dagger_\textbf{k} b_{\textbf{k}',\sigma} +  g_{\textbf{k}',\sigma}^*  b_{\textbf{k}',\sigma}^\dagger a_\textbf{k} ) \nonumber \\
    &&+ \sum_{\omega,d}   ( \kappa_{\omega,d}  s_{\omega,d}^\dagger a_\textbf{k} + \kappa_{\omega,d}^* a_\textbf{k}^\dagger s_{\omega,d} \nonumber ) \\
    &&+ \sum_{\omega,d,\sigma}  ( \gamma_{\omega,d,\sigma} s_{\omega,d}^\dagger b_{\textbf{k},\sigma} + \gamma_{\omega,d,\sigma}^* b_{\textbf{k},\sigma}^\dagger s_{\omega,d} ). 
\end{eqnarray}
Here $\omega_\text{p}$ and $\omega_\text{x}$ denote the photonic and excitonic mode frequencies, respectively. The operator $a_\textbf{k}^\dagger$ creates a PhC mode with crystal momentum $\textbf{k}$, while $b_{\textbf{k}',\sigma}^\dagger$ creates a plane-wave exciton mode with in-plane momentum $\textbf{k}' = \textbf{k} + \textbf{G}_{mn}$, where $\textbf{G}_{mn}$ is the reciprocal lattice vector of the PhC labeled by integers $m$ and $n$.
The index $\sigma \in {x,y}$ labels the two orthogonal in-plane polarizations of the exciton modes.
The operator $s^\dagger_{\omega,d}$  creates a free-space mode with in-plane momentum $\textbf{k}$ and frequency $\omega$, where the index $d$ specifies both the polarization (two orthogonal states) and the propagation direction (upward or downward).
Owing to the strong periodic index modulation in the PhC, the optical field of the PhC mode $a^\dagger_\textbf{k}$ at the monolayer plane consists of a superposition of spatial Fourier components with in-plane momenta $\textbf{k}' = \textbf{k} + \textbf{G}_{mn}$ \cite{sakoda2005optical}.
Momentum conservation therefore dictates that the PhC mode $a^\dagger_\textbf{k}$ couples exclusively to exciton modes with these specific momenta $b^\dagger_{\textbf{k}'}$ with coupling amplitudes $g_{\textbf{k}',\sigma}$ determined by the corresponding Fourier components of the PhC mode (see Supplementary Material).
This is to be distinguished from exciton-photon coupling in optical structures with continuous in-plane translation symmetry (e.g., distributed Bragg reflector cavities), where the optical field at the monolayer plane contains a single Fourier component and hence couples only to the exciton modes with the same in-plane momentum $\textbf{k}$.
The parameters $\kappa_{\omega,d}$ and $\gamma_{\omega,d,\sigma}$ characterize the coupling of the photonic and excitonic modes to the radiation continuum, respectively. 
These parameters can be independently extracted from measurements of the bare PhC slab and TMD material.
Notably, below the diffraction limit, only the fundamental mode $b^\dagger_{\textbf{k},\sigma}$ within this ensemble lies inside the light cone and can therefore radiate into the far field; all higher-order exciton modes with $\textbf{G}_{mn} \neq 0$ lie outside the light cone and are nonradiative. 

We now show how coherent exciton-photon coupling leads to the formation of a collective exciton mode and governs its radiative behavior.
Within the exciton-photon subspace 
\begin{eqnarray}
H^{(\text{p,x})}_\textbf{k}&=&\hbar \omega_\text{p} a^\dagger_\textbf{k} a_\textbf{k} 
+ \hbar \omega_\text{x} \sum_{\textbf{k}',\sigma} b_{\textbf{k}',\sigma}^\dagger b_{\textbf{k}',\sigma}\nonumber \\
&&+ \sum_{\textbf{k}',\sigma} ( g_{\textbf{k}',\sigma} a^\dagger_\textbf{k} b_{\textbf{k}',\sigma} +  g_{\textbf{k}',\sigma}^*  b_{\textbf{k}',\sigma}^\dagger a_\textbf{k} )\nonumber ,
\end{eqnarray}
the coupling of a single photonic mode to an ensemble of plane-wave excitonic modes is analogous to Dicke superradiance in momentum space. 
This Hamiltonian can be block-diagonalized by a unitary rotation within the exciton subspace, which isolates a single collective (superradiant or bright) exciton mode that alone couples to the photonic mode 
\begin{eqnarray}
x^\dagger_\textbf{k} =  1/g_\textbf{k} \sum_{\textbf{k}',\sigma} g^*_{\textbf{k}',\sigma} b^\dagger_{\textbf{k}',\sigma}, 
\end{eqnarray}
where $g_\textbf{k}=\sqrt{\sum_{\textbf{k}',\sigma} |g_{\textbf{k}',\sigma}|^2}$ is the collective exciton-photon coupling strength (see Supplementary Material).
The spatial profile of $x^\dagger_\textbf{k}$ mirrors that of the optical field at the monolayer plane, since its expansion coefficients are precisely the Fourier components of the field.
The remaining mutually orthogonal exciton states remain degenerate and decoupled from the photonic mode and are therefore omitted from the subsequent analysis. Note that these exciton states may still couple to the radiation continuum and thus remain visible in optical measurements (see Supplementary Material).
Importantly, the collective exciton mode $x^\dagger_\textbf{k}$ couples to the radiation continuum solely through its fundamental plane-wave component $b^\dagger_{\textbf{k},\sigma}$. 
Its radiative coupling is therefore $\beta_{\omega,d} = \sum_\sigma \gamma_{\omega,d,\sigma} \frac{g_{\mathbf{k},\sigma}}{g_\textbf{k}} $, which is generally suppressed relative to the bare exciton radiation ($\gamma_{\omega,d,\sigma}$) and vanishes entirely when the fundamental component is absent (i.e., $g_{\mathbf{k},\sigma}=0$).

In this rotated basis, the full Hamiltonian reduces to
\begin{eqnarray}
H_\textbf{k}&=&\hbar \omega_\text{p} a^\dagger_\textbf{k} a_\textbf{k} 
+ \hbar \omega_\text{x} x_\textbf{k}^\dagger x_\textbf{k} + \sum_{\omega,d}   \hbar \omega s^\dagger_{\omega,d} s_{\omega,d}  \nonumber \\
&&+ g_\textbf{k} a^\dagger_\textbf{k} x_\textbf{k} +  g_\textbf{k}  x_\textbf{k}^\dagger a_\textbf{k}  \nonumber \\
 &&+ \sum_{\omega,d}   ( \kappa_{\omega,d}  s_{\omega,d}^\dagger a_\textbf{k} + \kappa_{\omega,d}^* a_\textbf{k}^\dagger s_{\omega,d} ) \nonumber \\
 &&+ \sum_{\omega,d}   ( \beta_{\omega,d}  s_{\omega,d}^\dagger x_\textbf{k} + \beta_{\omega,d}^* x_\textbf{k}^\dagger s_{\omega,d} ). 
\end{eqnarray}
To proceed, we rewrite the Hamiltonian in the polariton basis defined by  
\begin{equation}
    \begin{split}
    p^\dagger_{\textbf{k},+} = c_{\text{x}} a^\dagger_\textbf{k} + c_{\text{p}} x^\dagger_\textbf{k} \\
    p^\dagger_{\textbf{k},-} = c_{\text{p}} a^\dagger_\textbf{k} - c_{\text{x}} x^\dagger_\textbf{k}
    \end{split}.
\end{equation}
Here $p^\dagger_{\textbf{k},\pm}$ are the creation operators for the upper (UP) and lower (LP) polariton branches. $c_{\text{p}}$ and $c_{\text{x}}$ denote the photonic and excitonic Hopfield coefficients of the lower polariton branch. In this basis, the Hamiltonian takes the form 
\begin{eqnarray}
H_\textbf{k}&=&\hbar \omega_{-} p^\dagger_{\textbf{k},-} p_{\textbf{k},-} + \hbar \omega_{+} p^\dagger_{\textbf{k},+} p_{\textbf{k},+} 
+ \sum_{\omega,d}   \hbar \omega s^\dagger_{\omega,d} s_{\omega,d}  \nonumber \\
    &&+ \sum_{\omega,d}  ( \tau_{\omega,d,-} s_{\omega,d}^\dagger p_{\textbf{k},-} + \tau_{\omega,d,-}^* p_{\textbf{k},-}^\dagger s_{\omega,d} \nonumber ) \\
    &&+ \sum_{\omega,d}  ( \tau_{\omega,d,+} s_{\omega,d}^\dagger p_{\textbf{k},+} + \tau_{\omega,d,+}^* p_{\textbf{k},+}^\dagger s_{\omega,d} ) \ \text{.}
\end{eqnarray}
Here, $\omega_\pm$ denote the UP and LP frequencies. The coefficients
\begin{equation}
\begin{split}
\tau_{\omega,d,+} &= c_\text{x} \kappa_{\omega,d} + c_\text{p} \beta_{\omega,d} \\
\tau_{\omega,d,-} &= c_\text{p} \kappa_{\omega,d} - c_\text{x} \beta_{\omega,d}
\end{split}
\label{eq:tau_pm}
\end{equation}
characterize the coupling of the polariton modes $p^\dagger_{\textbf{k},\pm}$ to the radiation continuum. 
Equation~\eqref{eq:tau_pm} reveals that polariton radiation arises from coherent interference between the photonic and collective exciton radiation channels, rather than from a weighted average of the bare photon and exciton radiation rates assumed in the coupled oscillator model. 

Building on this formalism, we illustrate how these interference mechanisms can dramatically suppress polariton radiation and give rise to polaritonic BICs in two realistic structures.
We first show that both collective exciton  and photonic radiation can vanish simultaneously at high-symmetry points in momentum space due to symmetry constraints, leading to symmetry-protected polaritonic BICs, as observed in various nanophotonic systems \cite{dang2022realization,kravtsov2020nonlinear,ardizzone2022polariton,wu2024exciton}.
Specifically, we consider a photonic band that is non-degenerate at the $\Gamma$ point.
The corresponding photonic mode at $\Gamma$ is even under the in-plane two-fold rotation ($C_2^z$) operation, whereas the free-space radiation modes are odd. As a result, the photonic mode cannot couple to the radiation continuum and forms a symmetry-protected photonic BIC \cite{hsu2016bound}.
When a TMD monolayer is strongly coupled to this photonic mode, the resulting collective exciton mode inherits the same symmetry. 
In contrast, the fundamental plane-wave exciton mode $b_{\Gamma,\sigma}^\dagger$ is odd under $C_2^z$, so it cannot contribute to the collective exciton mode at $\Gamma$. 
Consequently, the collective exciton mode is also nonradiative ($\beta_{\omega,d}=0$), giving rise to symmetry-protected polaritonic BICs at $\Gamma$.

\begin{figure*}[htbp]
\centering
\includegraphics[width=2\columnwidth]{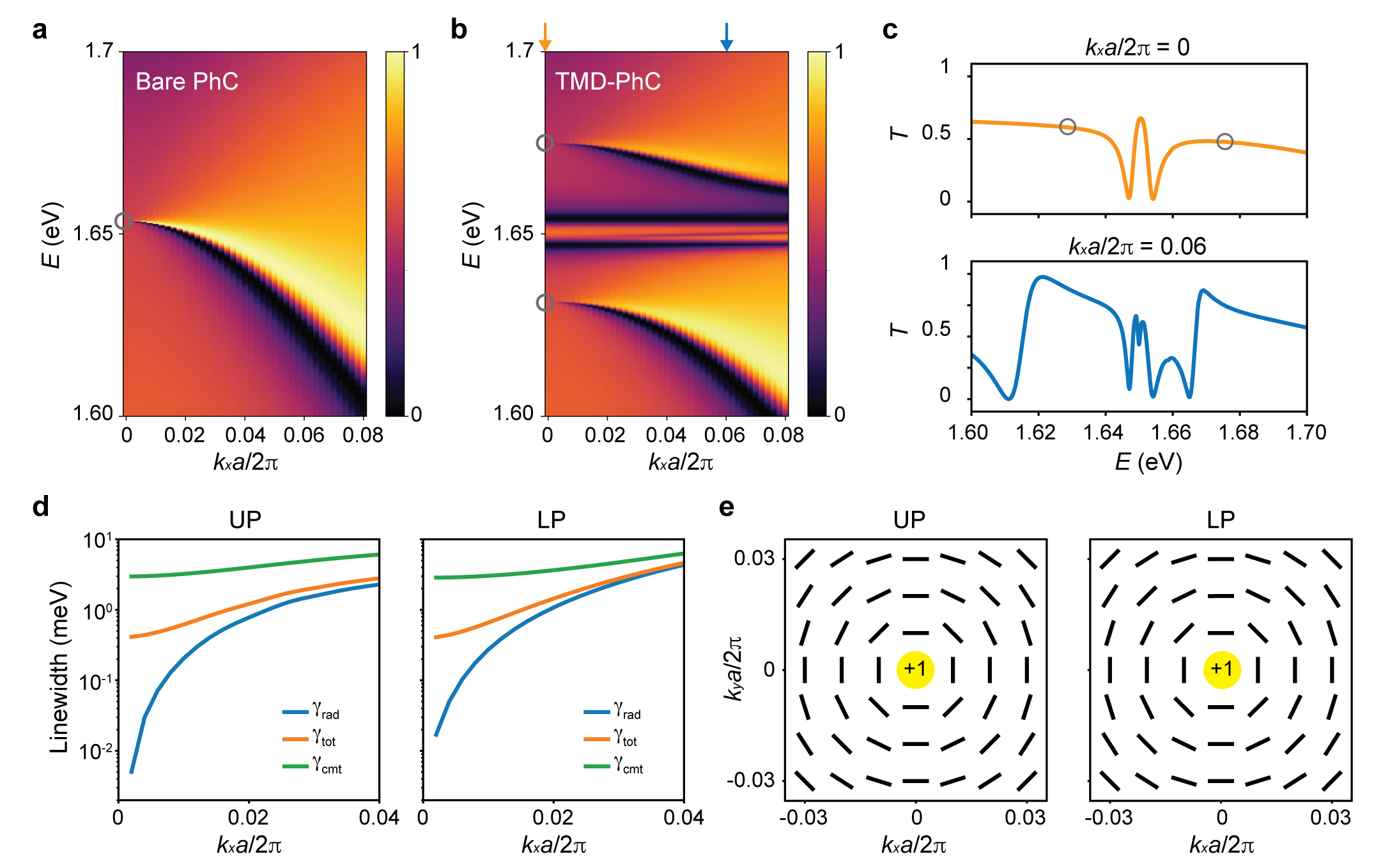}
\caption{\label{fig:pdf2} Symmetry-protected polaritonic BICs. a. Transmission spectra of the bare PhC slab under y-polarized excitation, showing an isolated photonic band that hosts a BIC at the $\Gamma$ point. b. Transmission spectra of the coupled TMD-PhC slab, showing two symmetry-protected polaritonic BICs at the $\Gamma$ point (highlighted by gray circles). c. Line cuts of transmission spectra at $k_x a/2\pi = 0$ (orange) and 0.06 (blue). d. Extracted total (orange) and radiative (blue) polariton linewidths, alongside the predicted total linewidths (green) from the coupled oscillator model. e. Far-field polarization map in momentum space showing vortex-like singularities, characteristic of polaritonic BICs.}
\end{figure*}

To verify this, we calculate the momentum-resolved transmission spectra of the coupled TMD-PhC slab using finite-difference time-domain (FDTD) simulations \cite{tidy3d}.
The PhC consists of a Si$_3$N$_4$ slab with thickness $t=150$ nm, lattice constant $a=550$ nm, and air-hole radius $r=195$ nm.
The TMD monolayer is placed 10 nm above the PhC slab and modeled as a Lorentz medium, with the exciton oscillator strength and nonradiative damping chosen such that it exhibits radiative and nonradiative linewidths of 5 meV and 0.8 meV, respectively, in vacuum \cite{zhou2020controlling,wang2025lorentz}.
As shown in Fig.~2a, the bare photonic BIC is nearly resonant with the exciton energy ($E_\text{x}=$ 1.654 eV).
When the TMD monolayer is introduced, strong exciton-photon hybridization occurs near the $\Gamma$ point, giving rise to upper and lower polariton branches (Fig.~2b). 
The resulting Rabi splitting of $\sim$40 meV agrees well with experimental values reported in similar TMD–PhC systems \cite{zhang2018photonic, wang2026strongly}, providing a consistency check for the simulation parameters.
In addition to the polaritonic bands, a nearly flat exciton band emerges within the polariton gap, corresponding to subradiant exciton modes that remain decoupled from the photonic bands. 
In contrast to the polaritonic BIC modes, these exciton modes are odd under $C^z_2$ and therefore remain radiatively visible at the $\Gamma$ point.
Notably, this flat exciton band exhibits additional fine spectral features, likely arising from weak hybridization with far-detuned photonic bands.  

To quantify the suppression of polariton radiation, we extract the radiative and total linewidths of the polariton modes by fitting the transmission spectra. 
As shown in Fig.~2d, the radiative linewidths (full width at half maximum) of both the LP and UP branches decrease monotonically as $\textbf{k}$ approaches $\Gamma$, ultimately vanishing. 
In contrast, their total linewidths saturate at half the nonradiative exciton linewidth ($\sim$0.4 meV), indicating that nonradiative exciton decay becomes the sole dissipation channel for the polaritonic BICs.
These linewidths stand in stark contrast to the predictions of the coupled oscillator model (green line in Fig.~2d), in which the polariton linewidths are given by a weighted average of the bare excitonic and photonic linewidths.
Furthermore, akin to their photonic counterparts, these polaritonic BICs exhibit a topological character, manifested by polarization vortices centered at $\Gamma$ in the far-field radiation pattern (Fig.~2e) \cite{zhen2014topological}. This topological character imparts robustness against structural perturbations that preserve $C_2^z$ symmetry. 

\begin{figure*}[htbp]
\includegraphics[width=2\columnwidth]{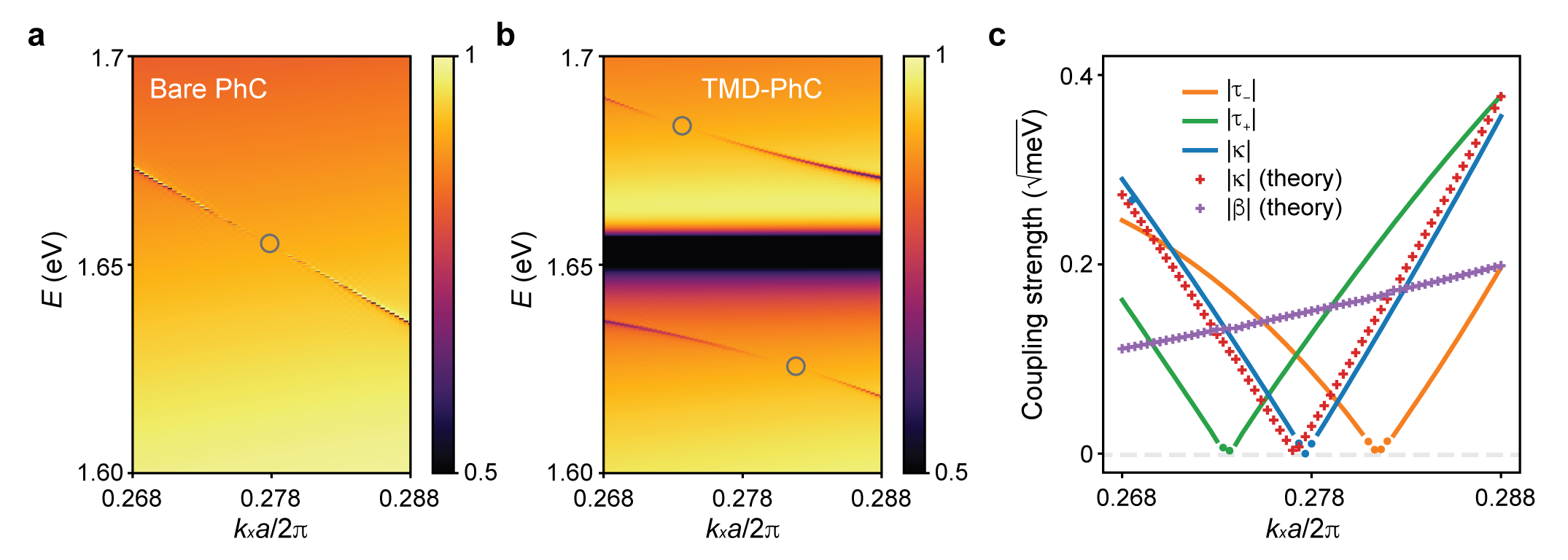}
\caption{\label{fig:pdf3}Interference-induced polaritonic BICs. a. Transmission spectra of the bare PhC slab under y-polarized excitation, showing a photonic BIC at $k_0 a/2\pi \approx 0.278$. b. Transmission spectra of the coupled TMD-PhC slab, showing two polaritonic BICs (highlighted by gray circles) shifted to either side of $k_0$. c. Radiative coupling strengths as a function of $k_x$: the measured polaritonic couplings $|\tau_{\pm}|$ (from the LP and UP linewidths) and the independently measured photonic coupling $|\kappa|$ (from the bare PhC linewidth), compared with values derived analytically. Because the coupling strengths vanish near the BICs and can not be determined precisely using FDTD simulations, these values are instead extrapolated as a linear function of $k_x$ (dots).}
\end{figure*}

In addition to symmetry-protected polaritonic BICs, another class can emerge at generic momentum points through destructive interference between the excitonic and photonic radiation channels. 
To demonstrate this mechanism explicitly, we consider a TMD monolayer embedded at the central plane of a PhC slab with thickness $t=470$ nm, lattice constant $a=330$ nm, and air-hole radius $r=97$ nm, such that the structure preserves up-down mirror symmetry ($\sigma_z$).
In the absence of the monolayer, the bare PhC slab hosts an off-$\Gamma$ photonic BIC at $k_0 a/2\pi \approx 0.278$, with its energy tuned to coincide with the bare exciton resonance (Fig.~3a).
When exciton-photon coupling is introduced, strong hybridization near $k_0$ gives rise to upper and lower polariton branches, each hosting a polaritonic BIC displaced to opposite sides of $k_0$ (Fig.~3b).
The emergence of these polaritonic BICs marks a significant departure from the coupled oscillator model, which predicts finite polariton radiation near $k_0$ because both the excitonic and photonic components remain radiative.

The emergence of these polaritonic BICs can be understood intuitively from the momentum dependence of $\kappa_{\omega,d}$ and $\beta_{\omega,d}$. 
Specifically, owing to the $\sigma_z$ mirror symmetry of the structure, the radiative coupling coefficients $\kappa_{\omega,d}$ and $\beta_{\omega,d}$ can be chosen to be real-valued along the $k_x$ axis under an appropriate gauge choice for the photonic and excitonic modes. 
Near the photonic BIC, $\kappa_{\omega,d}$ varies linearly with $k_x$, vanishing at $k_0$ and changing sign across it \cite{jin2019topologically}. 
In contrast, $\beta_{\omega,d}$ varies only weakly over the same momentum range. 
This allows $k_x$ to be tuned such that the excitonic and photonic radiation amplitudes cancel exactly at a momentum near $k_0$, giving rise to an off-$\Gamma$ polaritonic BIC.
Furthermore, because the relative phase between the collective exciton and photon modes differs by $\pi$ between the UP and LP branches, this cancellation occurs on opposite sides of $k_0$ for the two branches, consistent with Fig.~3b. 

\begin{figure*}[htbp]
\includegraphics[width=2\columnwidth]{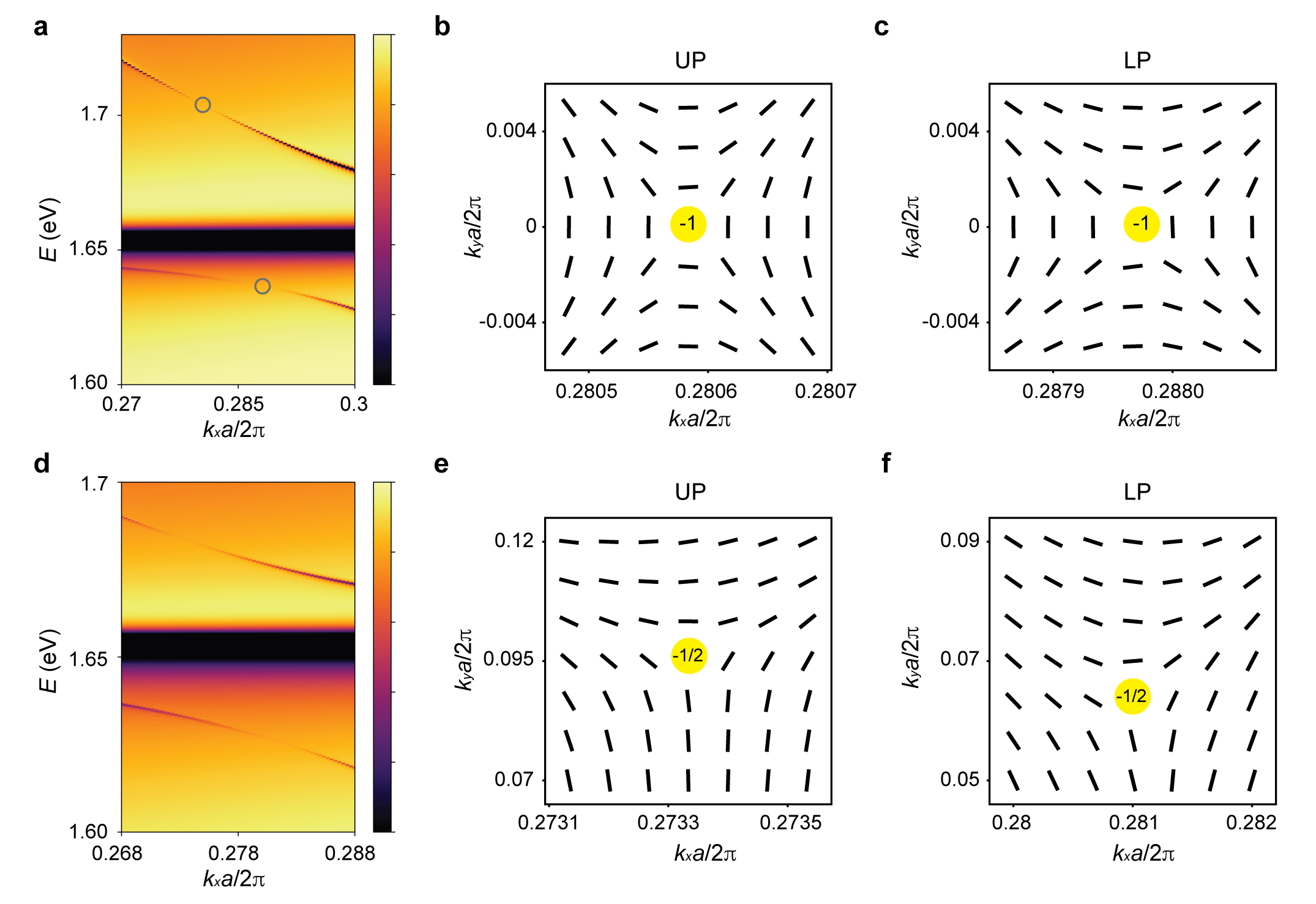}
\caption{\label{fig:pdf4} Robustness of off-$\Gamma$ polaritonic BICs. a. Transmission spectra of the coupled TMD-PhC slab after enlarging the PhC hole radius from 97 nm to 105 nm (a $\sigma_z$-preserving perturbation), showing that the polaritonic BICs persist but shift asymmetrically. b,c. Far-field polarization maps for the UP and LP branches in (a), showing taht each BIC retains an integer topological charge. d. Transmission spectra showing finite polariton linewidths when the TMD monolayer is displaced 5 nm away from the central plane (breaking $\sigma_z$, indicating the breakdown of the polaritonic BICs. e,f. Far-field polarization maps for the UP and LP branches in (d), showing that the original integer polarization vortex splits into a pair of half-integer vortices with the same winding number.}
\end{figure*}

To verify this picture quantitatively, we extract the relevant radiative coupling coefficients directly from the simulated transmission spectra.
Specifically, we introduce the time-domain radiative coupling coefficients $\kappa$, $\beta$, and $\tau_\pm$, whose squared magnitudes are the corresponding radiative decay rates.
The photonic coupling $\kappa$ is therefore obtained from the linewidth of the bare PhC slab, while the polaritonic radiative couplings $\tau_{\pm}$ are obtained from the linewidths of the UP and LP branches (to isolate the polariton radiative linewidths, we set the exciton nonradiative damping to zero in these simulations). 
These time-domain coupling coefficients are related to the frequency-domain coefficients $\kappa_{\omega,d}$, $\beta_{\omega,d}$, and $\tau_{\omega,d,\pm}$ introduced in our Hamiltonian. For the bands along the $k_x$ axis, both the photonic and collective exciton radiation are $y$-polarized, while the upward and downward radiation channels are related by $\sigma_z$ symmetry, allowing us to suppress the radiation-channel index $d$. Furthermore, because the frequency-domain coupling coefficients vary negligibly over the narrow spectral range considered here, they are proportional to their corresponding time-domain coefficients up to a constant normalization factor.
As shown in Fig.~3c, $\kappa$ vanishes linearly at the bare photonic BIC $k_0$, while $\tau_{\pm}$ each vanish linearly at their respective polaritonic BIC momenta, consistent with the picture above.
However, the excitonic radiative coupling $\beta$ cannot be extracted directly from the spectra. 
Instead, we derive both $\kappa$ and $\beta$ from the extracted $\tau_{\pm}$ and the Hopfield coefficients by inverting Eq.~\eqref{eq:tau_pm}. 
As shown in Fig.~3c, the derived $\kappa$ closely matches the value independently extracted from the bare PhC slab, validating our procedure. 
The slight remaining discrepancy likely arises from weak hybridization with additional, far-detuned photonic bands not captured by our two-band coupling model.
Meanwhile, the derived $\beta$ varies only weakly across this momentum range, consistent with its physical origin as the projection of the collective exciton mode $x^\dagger_\textbf{k}$ onto the fundamental exciton mode $b^\dagger_{\textbf{k},\sigma}$.
The agreement between the derived and independently extracted $\kappa$, together with the weak momentum dependence of $\beta$, confirms that the off-$\Gamma$ polaritonic BICs arise directly from destructive interference between the excitonic and photonic radiation channels.

Like their symmetry-protected counterparts, these off-$\Gamma$ polaritonic BICs exhibit polarization vortices in their far-field radiation, indicating their nontrivial topological character. 
This topological character renders them robust against structural perturbations that preserve $\sigma_z$ symmetry. 
To illustrate this, we perturb the design by enlarging the PhC air-hole radius from 97 nm to 105 nm, which blue-shifts the photonic BIC energy while leaving the exciton resonance unchanged.
As shown in Fig.~4a, the upper and lower polaritonic BICs persist under this detuning, but their momentum-space locations shift asymmetrically about $k_0$ -- in contrast to the symmetric placement observed in the resonant case (Fig.~3b). 
The corresponding far-field polarization maps (Fig.~4b,c) show that the perturbation merely shifts the integer topological charges in momentum space without destroying them.   
This demonstrates that the off-$\Gamma$ polaritonic BICs are a generic phenomenon rather than a consequence of fine-tuned parameters.

In contrast, this protection relies critically on $\sigma_z$ symmetry: perfect destructive interference breaks down when $\sigma_z$ symmetry is broken. To demonstrate this, we simulate the transmission spectra of a structure where the TMD monolayer is displaced by 5 nm from the central plane. As shown in Fig. 4d, both the upper and lower polaritonic BICs disappear, and the polariton linewidths remain finite near $k_0$. This behavior arises because both the collective excitonic and photonic radiative couplings become generically complex, making tuning $k_x$ alone insufficient to achieve perfect destructive interference. This breakdown can also be understood from a topological perspective. As shown in Fig. 4e,f, breaking $\sigma_z$ causes the integer polarization vortex, originally pinned to the $k_x$ axis and associated with vanishing far-field radiation, to split into a pair of half-integer vortices with the same winding number. These vortices are displaced symmetrically about the $k_x$-axis and exhibit finite far-field radiation. A similar phenomenon has been observed in dielectric photonic crystals with broken up-down mirror symmetry \cite{yin2020observation}.

In summary, we present a theoretical framework for investigating the radiative decay of exciton-polaritons in nanophotonic structures that goes beyond the conventional coupled oscillator model.
We demonstrate that the collective and coherent nature of exciton-photon interactions can dramatically modify polariton radiation in nanophotonic environments with spatially varying optical fields, enabling the formation of polaritonic BICs with theoretically infinite radiative lifetimes. 
While our analysis focuses on the coupling between a TMD monolayer and a periodically patterned PhC slab in momentum space, the underlying physics -- collective exciton mode formation and interference-based radiative suppression -- does not itself rely on periodicity.
Therefore, our framework extends naturally to nanophotonic structures without translation symmetry, such as photonic crystal nanocavities \cite{wang2026strongly} and plasmonic structures \cite{liu2016strong}.
Likewise, the underlying principles are broadly applicable to other excitonic material systems, such as III-V quantum wells \cite{deng2010exciton} and perovskites.
This generality provides a broad foundation for engineering long-lived exciton-polaritons across diverse material platforms, offering promising prospects for both classical and quantum polaritonic applications, including low-threshold nonlinear polaritonics \cite{wang2026strongly}, topological polaritonics \cite{he2023polaritonic}, and quantum polaritonic phenomena \cite{delteil2019towards,munoz2019emergence}.
\\

\section*{Author Contributions}
L.H. and B.Z. conceived the idea. J.S. and L.H. developed the theory and performed the numerical simulations. All authors discussed the results and wrote the paper. 

\begin{acknowledgments}
This work was partly supported by the U.S. Office of Naval Research (ONR) through grant N00014-20-1-2325 on Robust Photonic Materials with High-Order Topological Protection and grant N00014-21-1-2703, as well as the Sloan Foundation. L.H. acknowledges support from the MonArk NSF Quantum Foundry supported by the National Science Foundation Q-AMASE-i program under NSF Award No. DMR-1906383.
\end{acknowledgments}

\bibliography{apssamp}

@article{wang2025lorentz,
  title={Lorentz--Drude Dipoles in the Radiative Limit and Their Modeling in Finite-Difference Time-Domain Methods},
  author={Wang, Heming and Fan, Shanhui},
  journal={Annalen der Physik},
  pages={e00156},
  year={2025},
  publisher={Wiley Online Library}
}

@article{zhou2020controlling,
  title={Controlling excitons in an atomically thin membrane with a mirror},
  author={Zhou, You and Scuri, Giovanni and Sung, Jiho and Gelly, Ryan J and Wild, Dominik S and De Greve, Kristiaan and Joe, Andrew Y and Taniguchi, Takashi and Watanabe, Kenji and Kim, Philip and others},
  journal={Physical review letters},
  volume={124},
  number={2},
  pages={027401},
  year={2020},
  publisher={APS}
}

@article{yin2020observation,
  title={Observation of topologically enabled unidirectional guided resonances},
  author={Yin, Xuefan and Jin, Jicheng and Solja{\v{c}}i{\'c}, Marin and Peng, Chao and Zhen, Bo},
  journal={Nature},
  volume={580},
  number={7804},
  pages={467--471},
  year={2020},
  publisher={Nature Publishing Group UK London}
}

@article{zhen2014topological,
  title={Topological nature of optical bound states in the continuum},
  author={Zhen, Bo and Hsu, Chia Wei and Lu, Ling and Stone, A Douglas and Solja{\v{c}}i{\'c}, Marin},
  journal={Physical review letters},
  volume={113},
  number={25},
  pages={257401},
  year={2014},
  publisher={APS}
}

@article{hsu2016bound,
  title={Bound states in the continuum},
  author={Hsu, Chia Wei and Zhen, Bo and Stone, A Douglas and Joannopoulos, John D and Solja{\v{c}}i{\'c}, Marin},
  journal={Nature Reviews Materials},
  volume={1},
  number={9},
  pages={1--13},
  year={2016},
  publisher={Nature Publishing Group}
}

@article{dang2022realization,
  title={Realization of polaritonic topological charge at room temperature using polariton bound states in the continuum from perovskite metasurface},
  author={Dang, Nguyen Ha My and Zanotti, Simone and Drouard, Emmanuel and Chevalier, C{\'e}line and Tripp{\'e}-Allard, Ga{\"e}lle and Amara, Mohamed and Deleporte, Emmanuelle and Ardizzone, Vincenzo and Sanvitto, Daniele and Andreani, Lucio Claudio and others},
  journal={Advanced Optical Materials},
  volume={10},
  number={6},
  pages={2102386},
  year={2022},
  publisher={Wiley Online Library}
}

@article{wu2024exciton,
  title={Exciton polariton condensation from bound states in the continuum at room temperature},
  author={Wu, Xianxin and Zhang, Shuai and Song, Jiepeng and Deng, Xinyi and Du, Wenna and Zeng, Xin and Zhang, Yuyang and Zhang, Zhiyong and Chen, Yuzhong and Wang, Yubin and others},
  journal={Nature Communications},
  volume={15},
  number={1},
  pages={3345},
  year={2024},
  publisher={Nature Publishing Group UK London}
}

@article{kravtsov2020nonlinear,
  title={Nonlinear polaritons in a monolayer semiconductor coupled to optical bound states in the continuum},
  author={Kravtsov, Vasily and Khestanova, Ekaterina and Benimetskiy, Fedor A and Ivanova, Tatiana and Samusev, Anton K and Sinev, Ivan S and Pidgayko, Dmitry and Mozharov, Alexey M and Mukhin, Ivan S and Lozhkin, Maksim S and others},
  journal={Light: Science \& Applications},
  volume={9},
  number={1},
  pages={56},
  year={2020},
  publisher={Nature Publishing Group UK London}
}

@article{ardizzone2022polariton,
  title={Polariton Bose--Einstein condensate from a bound state in the continuum},
  author={Ardizzone, V and Riminucci, F and Zanotti, S and Gianfrate, A and Efthymiou-Tsironi, M and Su{\`a}rez-Forero, DG and Todisco, F and De Giorgi, M and Trypogeorgos, D and Gigli, G and others},
  journal={Nature},
  volume={605},
  number={7910},
  pages={447--452},
  year={2022},
  publisher={Nature Publishing Group UK London}
}

@article{he2023polaritonic,
  title={Polaritonic Chern insulators in monolayer semiconductors},
  author={He, Li and Wu, Jingda and Jin, Jicheng and Mele, Eugene J and Zhen, Bo},
  journal={Physical Review Letters},
  volume={130},
  number={4},
  pages={043801},
  year={2023},
  publisher={APS}
}

@article{zhang2018photonic,
  title={Photonic-crystal exciton-polaritons in monolayer semiconductors},
  author={Zhang, Long and Gogna, Rahul and Burg, Will and Tutuc, Emanuel and Deng, Hui},
  journal={Nature communications},
  volume={9},
  number={1},
  pages={713},
  year={2018},
  publisher={Nature Publishing Group UK London}
}

@article{byrnes2014exciton,
  title={Exciton--polariton condensates},
  author={Byrnes, Tim and Kim, Na Young and Yamamoto, Yoshihisa},
  journal={Nature Physics},
  volume={10},
  number={11},
  pages={803--813},
  year={2014},
  publisher={Nature Publishing Group UK London}
}

@article{zhang2021van,
  title={Van der Waals heterostructure polaritons with moir{\'e}-induced nonlinearity},
  author={Zhang, Long and Wu, Fengcheng and Hou, Shaocong and Zhang, Zhe and Chou, Yu-Hsun and Watanabe, Kenji and Taniguchi, Takashi and Forrest, Stephen R and Deng, Hui},
  journal={Nature},
  volume={591},
  number={7848},
  pages={61--65},
  year={2021},
  publisher={Nature Publishing Group UK London}
}

@book{sakoda2005optical,
  title={Optical properties of photonic crystals},
  author={Sakoda, Kazuaki},
  year={2005},
  publisher={Springer}
}

@misc{tidy3d,
  author = {{Flexcompute Inc.}},
  title = {Tidy3D: Spatiotemporal electromagnetic solver},
  year = {2025},
  howpublished = {\url{https://www.flexcompute.com/tidy3d/}},
  note = {}
}

@article{chen2020metasurface,
  title={Metasurface integrated monolayer exciton polariton},
  author={Chen, Yueyang and Miao, Shengnan and Wang, Tianmeng and Zhong, Ding and Saxena, Abhi and Chow, Colin and Whitehead, James and Gerace, Dario and Xu, Xiaodong and Shi, Su-Fei and others},
  journal={Nano Letters},
  volume={20},
  number={7},
  pages={5292--5300},
  year={2020},
  publisher={ACS Publications}
}

@article{khestanova2024electrostatic,
  title={Electrostatic control of nonlinear photonic-crystal polaritons in a monolayer semiconductor},
  author={Khestanova, Ekaterina and Shahnazaryan, Vanik and Kozin, Valerii K and Kondratyev, Valeriy I and Krizhanovskii, Dmitry N and Skolnick, Maurice S and Shelykh, Ivan A and Iorsh, Ivan V and Kravtsov, Vasily},
  journal={Nano Letters},
  volume={24},
  number={24},
  pages={7350--7357},
  year={2024},
  publisher={ACS Publications}
}

@article{deng2010exciton,
  title={Exciton-polariton bose-einstein condensation},
  author={Deng, Hui and Haug, Hartmut and Yamamoto, Yoshihisa},
  journal={Reviews of modern physics},
  volume={82},
  number={2},
  pages={1489--1537},
  year={2010},
  publisher={APS}
}

@article{dicke1954coherence,
  title={Coherence in spontaneous radiation processes},
  author={Dicke, Robert H},
  journal={Physical review},
  volume={93},
  number={1},
  pages={99},
  year={1954},
  publisher={APS}
}

@article{sanvitto2016road,
  title={The road towards polaritonic devices},
  author={Sanvitto, Daniele and K{\'e}na-Cohen, St{\'e}phane},
  journal={Nature materials},
  volume={15},
  number={10},
  pages={1061--1073},
  year={2016},
  publisher={Nature Publishing Group UK London}
}

@article{munoz2019emergence,
  title={Emergence of quantum correlations from interacting fibre-cavity polaritons},
  author={Mu{\~n}oz-Matutano, Guillermo and Wood, Andrew and Johnsson, Mattias and Vidal, Xavier and Baragiola, Ben Q and Reinhard, Andreas and Lema{\^\i}tre, Aristide and Bloch, Jacqueline and Amo, Alberto and Nogues, Gilles and others},
  journal={Nature materials},
  volume={18},
  number={3},
  pages={213--218},
  year={2019},
  publisher={Nature Publishing Group UK London}
}

@article{delteil2019towards,
  title={Towards polariton blockade of confined exciton--polaritons},
  author={Delteil, Aymeric and Fink, Thomas and Schade, Anne and H{\"o}fling, Sven and Schneider, Christian and {\.I}mamo{\u{g}}lu, Ata{\c{c}}},
  journal={Nature materials},
  volume={18},
  number={3},
  pages={219--222},
  year={2019},
  publisher={Nature Publishing Group UK London}
}

@article{liu2016strong,
  title={Strong exciton--plasmon coupling in MoS2 coupled with plasmonic lattice},
  author={Liu, Wenjing and Lee, Bumsu and Naylor, Carl H and Ee, Ho-Seok and Park, Joohee and Johnson, AT Charlie and Agarwal, Ritesh},
  journal={Nano letters},
  volume={16},
  number={2},
  pages={1262--1269},
  year={2016},
  publisher={ACS Publications}
}

@article{wang2018colloquium,
  title={Colloquium: Excitons in atomically thin transition metal dichalcogenides},
  author={Wang, Gang and Chernikov, Alexey and Glazov, Mikhail M and Heinz, Tony F and Marie, Xavier and Amand, Thierry and Urbaszek, Bernhard},
  journal={Reviews of Modern Physics},
  volume={90},
  number={2},
  pages={021001},
  year={2018},
  publisher={APS}
}

@article{zanotti2022theory,
  title={Theory of photonic crystal polaritons in periodically patterned multilayer waveguides},
  author={Zanotti, Simone and Nguyen, Hai Son and Minkov, Momchil and Andreani, Lucio C and Gerace, Dario},
  journal={Physical Review B},
  volume={106},
  number={11},
  pages={115424},
  year={2022},
  publisher={APS}
}

@article{zanotti2024legume,
  title={Legume: A free implementation of the guided-mode expansion method for photonic crystal slabs},
  author={Zanotti, Simone and Minkov, Momchil and Nigro, Davide and Gerace, Dario and Fan, Shanhui and Andreani, Lucio Claudio},
  journal={Computer Physics Communications},
  volume={304},
  pages={109286},
  year={2024},
  publisher={Elsevier}
}

@article{genco2025femtosecond,
  title={Femtosecond switching of strong light-matter interactions in microcavities with two-dimensional semiconductors},
  author={Genco, Armando and Louca, Charalambos and Cruciano, Cristina and Song, Kok Wee and Trovatello, Chiara and Di Blasio, Giuseppe and Sansone, Giacomo and Randerson, Sam A and Claronino, Peter and Georgiou, Kyriacos and others},
  journal={Nature communications},
  volume={16},
  number={1},
  pages={6490},
  year={2025},
  publisher={Nature Publishing Group UK London}
}

@article{wang2026strongly,
  title={Strongly Nonlinear Nanocavity Exciton Polaritons in Gate-Tunable Monolayer Semiconductors},
  author={Wang, Zhi and Kim, Bumho and Zhen, Bo and He, Li},
  journal={Physical Review Letters},
  volume={136},
  number={14},
  pages={146901},
  year={2026},
  publisher={APS}
}

@article{jin2019topologically,
  title={Topologically enabled ultrahigh-Q guided resonances robust to out-of-plane scattering},
  author={Jin, Jicheng and Yin, Xuefan and Ni, Liangfu and Solja{\v{c}}i{\'c}, Marin and Zhen, Bo and Peng, Chao},
  journal={Nature},
  volume={574},
  number={7779},
  pages={501--504},
  year={2019},
  publisher={Nature Publishing Group UK London}
}

\end{document}